\documentclass[a4paper, 10pt,]{IEEEtran}
\usepackage[utf8]{inputenc}
\usepackage{blindtext, graphicx}
\usepackage{listings}
\usepackage{graphicx}
\usepackage{floatrow}
\usepackage{amsmath}
\usepackage{mathtools}
\usepackage{esvect}
\usepackage{amsfonts}
\usepackage{booktabs}
\usepackage{siunitx}
\usepackage[thinlines]{easytable}
\usepackage{textcomp,gensymb}
\usepackage{epstopdf}
\usepackage{csquotes}
\usepackage[greek,english]{babel}
\usepackage{alphabeta}

\usepackage{array}

\usepackage{graphicx}
\usepackage[caption=false,font=small,style=base]{subfig}
\usepackage{etoolbox}
\makeatletter
\patchcmd{\@makecaption}
  {\scshape}
  {}
  {}
  {}
\makeatletter
\patchcmd{\@makecaption}
  {\\}
  {.\ }
  {}
  {}
\makeatother

\usepackage[style = numeric, sorting = none,url= false, doi = false, isbn = false]{biblatex}
\title{High contrast Ultrafast 3D Ultrasound Imaging using Row Column specific Frame Multiply and Sum}

\author{
\IEEEauthorblockN{Joseph Hansen-Shearer, Marcelo Lerendegui, Matthieu Toulemonde, Meng-Xing Tang}
}

\begin{document}

\maketitle
\thispagestyle{empty}
\pagestyle{empty}

\begin{abstract}

Row-column arrays have shown to be able to generate 3-D ultrafast ultrasound images with an order of magnitude less independent electronic channels than classic 2D matrix arrays. Unfortunately row-column array images suffer from major imaging artefacts due to the high side lobes. This paper proposes a row-column specific beamforming technique that exploits the incoherent nature of certain row column array artefacts. The geometric mean of the data from each row and column pair is taken prior to summation in beamforming, thus drastically reducing incoherent imaging artefacts compared to traditional coherent compounding. The effectiveness of this technique was demonstrated {\it{in silico}}, and the results show an average fivefold reduction in side-lobe levels. Significantly improved contrast was demonstrated with Tissue-to-noise ratio increasing from $\sim$10dB to $\sim$30dB and Tissue Contrast Ratio increasing from $\sim$21dB to $\sim$42dB  when using the proposed new method compared to Delay and Sum. These new techniques allowed for high quality 3D imaging whilst maintaining high frame rate potential.


\end{abstract}

\section{Introduction/Theory}

\begin{figure} [H]
    \centering
        \includegraphics[width=\linewidth]{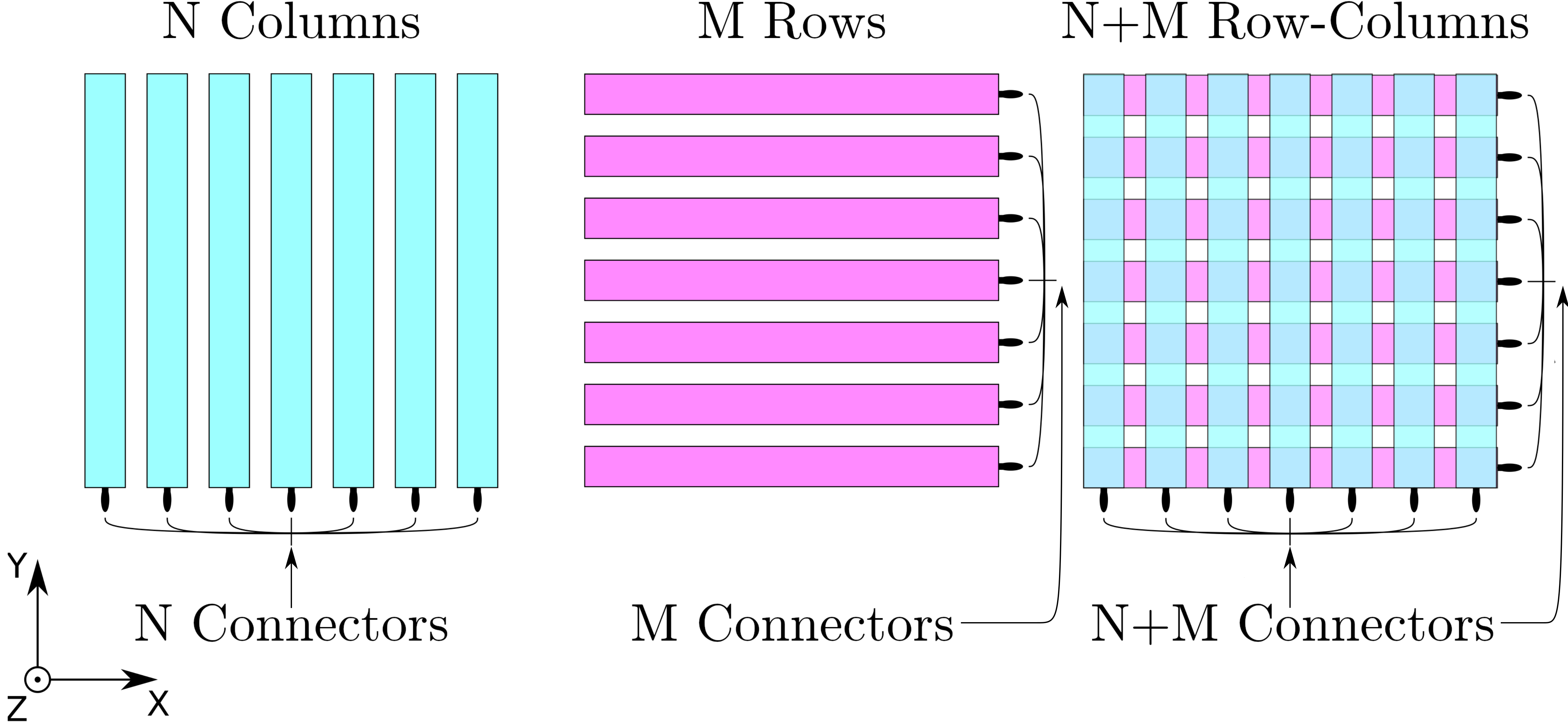}
        \caption{Diagram showing the basic configuration of the row-column array transducer. The probe has N columns and M rows. The rows are aligned along the X direction and the columns are aligned along the Y direction.}
        \label{fig:RCA_Diagram}
\end{figure}

Real-time 3-D ultrasound imaging is an emerging field in ultrasound research. Currently different techniques are used to produce 3D images of the body using ultrasound. Ideally a fully addressed 2-D matrix array probe would be used~\cite{gelly_two-dimensional_1998}. Unfortunately these probes require a prohibitively high number of electronic channels. If matrix probe has $N$ elements in the elevational direction and $M$ elements in the lateral direction, then for 
the matrix probe to be fully addressed it will require $N \times M$ elements and thus $N \times M$ electronic channels. Handling this quantity of channels is not feasible and thus alternative imaging techniques are required~\cite{huang_review_2017}. Many different approaches to solve 3-D imaging have been proposed and are currently being investigated, including sparsely addressed matrix probes~\cite{harput_3-d_2018}, multiplex matrix array probes~\cite{hara_new_2005} and mechanically stirred 1-D probes~\cite{fenster_three-dimensional_2001}. This paper will focus on a Row-Column Array (RCA) probe~\cite{morton_theoretical_2003}. 
An RCA probe consist of a set of elements aligned in one dimension (the rows) and the other set of elements are orthogonally aligned in a second dimension (the columns). All the elements are elongated in one dimension such that their footprint covers the entire probe surface, see Figure~\ref{fig:RCA_Diagram}. By using a RCA probe the number of channels can be reduced to $N+M$. This greatly decreases the hardware requirement and computational cost associated with 3-D imaging. The RCA can either transmit using the rows or with the columns. In reception again either the rows or the columns can be used. The RCA probe currently is operated using a slightly amended Delay and Sum (DAS) algorithm along with coherent compounding. The transmitted beamforms are either a full plane wave, whereby the entire volume is imaged during each transmission,~\cite{chen_column-row-parallel_2018} or focused onto a single plane~\cite{rasmussen_3-d_2015}, whereby multiple transmissions are required to reconstruct a full volume.  As this study is concentrating on fast 3-D ultrasound imaging only, the full volume reconstruction technique will be considered, as this is considerably faster than the plane-by-plane technique. Since RCA probes were fist introduced much work has been done to demonstrated the value of RCA probes~\cite{seo_5a-5_2006,demore_real-time_2009,sampaleanu_top-orthogonal--bottom-electrode_2014}, along with development of apodisation schemes to account for ghost echos and other artefacts artefacts~\cite{rasmussen_3-d_2015,christiansen_3-d_2015}, ultrafast RCA imaging for flow estimation~\cite{holbek_3-d_2016,flesch_4d_2017,sauvage_large_2018,sauvage_ultrafast_2018} and recently super resolution has been demonstrated using RCA probes~\cite{jensen_three-dimensional_2019}.

 Although RCA imaging is a promising technique for producing volumetric ultrasound images it suffers from considerable imaging artefacts. These artefacts, located in the lateral and elevation direction, are large and {\it{`cross-like'}} because each transmission event only has spatial information in two of the three dimensions. This means that the PSF produced from a single transmission-reception event will have a line shape rather than a circular shape (as is the case with traditional ultrasound). Thus to get complete information, an orthogonal transmission is needed for full spatial localisation. This will create a {\it{`cross-shaped'}} PSF.  Coherent compounding, using multiple transmission angles, can reduce  the {\it{`cross-shape’}} artefacts, but these are still prevalent particularly when fewer transmission angles are used.  This is an issue when performing fast ultrasound imaging as fewer transmissions are possible.

 Here we propose to generate high contrast, low artefact ultrafast 3D ultrasound using a coherence beamforming technique tailored specifically for RCA probes, which takes advantage of the incoherence of row and column transmissions. Coherence beamforming is an emerging technique that would be used as an alternative for DAS beamforming. Many different coherence beamforming techniques have been recently proposed including Delay Multiply and Sum~\cite{matrone_delay_2015}, Frame Multiply and Sum (FMAS)~\cite{kang_ultrafast_2020}, coherence factor~\cite{yang_united_2020, fatemi_rowcolumn-based_2020}, short-lag spatial coherence~\cite{lediju_short-lag_2011} and acoustic subaperture processing (ASAP)~\cite{stanziola_asap_2018} amongst others. We are not aware of any existing studies to apply coherent based beamforming to RCAs. This paper will focus on adapting the FMAS technique to RCA probes and also develop a new Row-Column specific Frame Multiply and Sum {RC-FMAS}. These two methods will be evaluated {\it{in silico}} and compared to traditional DAS beamforming. Additionally an optimisation sweep will be performed to evaluate the regimes where these new techniques are most effective. 

\section{Methods}
\subsection{In-phase Quadrature Demodulation}
The first step of this new beamforming technique is to perform In-phase Quadrature (IQ) Demodulation to the Radio Frequency (RF) data. When performing FMAS beamforming the multiplication of the signals will double any frequency component of the RF signals. As a consequence very high spatial sampling is required.
IQ Demodulation will shift the central frequency to zero, allowing  for the use of lower spatial sampling frequencies. The IQ demodulated data ($RF_{IQ}$) is then given by:
\begin{equation}
RF_{IQ}(n,t,θ) = RF(n,t,θ) e^{-i 2π f_c t },
\end{equation}
where $RF(n,t,θ) $ is the RF data of channel $n$, time interval t and plane wave transmission angle $θ$. 
In 3D imaging the angle will either be in the lateral or elevational direction.
After this a low-pass filter is then applied to $RF_{IQ}$ to only accept frequencies within a down mixed bandwidth. 

\subsection{Row-Column Delay and Sum}
To generate images using a RCA probe DAS beamforming can be performed. However, the delay algorithm needs to be amended to allow for elongated elements~\cite{flesch_4d_2017}. The forward delay ($d_{Tx}$) calculations are identical to standard plane wave calculations:

\begin{equation}
    d_{Tx}(x,z,θ) = \frac{z \cdot cos(θ) + x \cdot sin(θ)}{c},
\end{equation}

where $z$ is the depth, $θ$ is the angle of transmission and $c$ is the assumed speed of sound in the tissue. $x$ is either the lateral or elevational position depending on if transmission is with rows or columns. The back-propagation delay ($d_{Rx}$) is then calculated using the shortest distance to the corresponding row or column: 

\begin{equation}
    d_{Rx}(y,z,n) = \frac{\sqrt{z^{2}+(y-r(n))^{2}}}{c},
\end{equation}

where $y$ is either the elevational or lateral position and $r(n)$ is the position of the n\textsuperscript{th} row or column depending on which is being used for reception. The total delay $d_{tot}$ is then just given by 

\begin{equation}
    d_{tot}(n,R,θ) = d_{Tx}(x,z,θ)+d_{Rx}(y,z,r_{n}), 
\end{equation}

where $R$ is the position of voxel located at $(x,y,z)$. In addition to the total delay phase compensation is needed to account for IQ Demodulation. After calculating the delay on each voxel a volume is generated from each angled transmission by taking the corresponding signal from each channel and summing them together. The resulting volume $RF_{DAS}$ is given by 

\begin{equation}
RF_{DAS}(R,θ) = \sum_{n=1}^{N} RF_{IQ} (n,d_{tot}(n,R,θ),θ),
\end{equation}
where N is either the number of rows or columns depending on which is being used in transmission. 
When performing traditional coherent compounding multiple transmission would be sent at different angles, and then a volume would be generated from this. Figure~\ref{fig:DAS_Diagram} shows pipeline used to generate row column delay and sum images. In FMAS a different procedure, outlined below, will be followed.

\subsection{Frame Multiply and Sum}
When performing FMAS each $RF_{DAS}$ volume, i.e. a 3D image reconstructed using a single plane wave transmission, is coupled with all possible other volumes. The resulting number of pairs $N_P$ is thus given by

\begin{equation}
N_P = \binom{N_{Tx}}{2} = \frac{N_{Tx}^2-N_{Tx}}{2},
\end{equation}

where $N_{Tx}$ here is the number of plane wave transmissions.
The paired volumes are then combined by taking the {\it{'signed'}} geometric mean of the paired voxels. These combined pairs are then summed to give a resulting volume.
If the product of the pairs $i$ and $j$  ($RF_{ij}$) is given by

\begin{equation}
RF_{ij} = RF_{DAS}(R,i)RF_{DAS}(R,j),
\end{equation}

then the resulting delayed multiplied and summed volume $RF_{FMAS}$ will be

\begin{equation}
RF_{FMAS} = \sum_{i= 1}^{N_{Tx}-1} \sum_{j=i+1}^{N_{Tx}} sign(RF_{ij}) \cdot \sqrt{\Big\vert RF_{ij}\Big\vert}.
\end{equation}

The final volume will then be produced by taking the absolute value of this sum and then log compressing. Figure~\ref{fig:FMAS_pipeline} shows a schematic representation of the FMAS pipeline 

\subsection{Row-Column specific Frame Multiply and Sum}

\begin{figure}[H]
\centering

\subfloat[Traditional Delay and Sum Pipeline \label{fig:DAS_Diagram}]{\includegraphics[width=\linewidth]{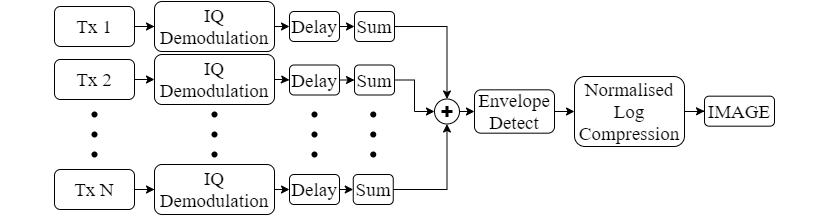}} \hfill
\subfloat[Frame Multiply and Sum Pipeline \label{fig:FMAS_pipeline}]{\includegraphics[width=\linewidth]{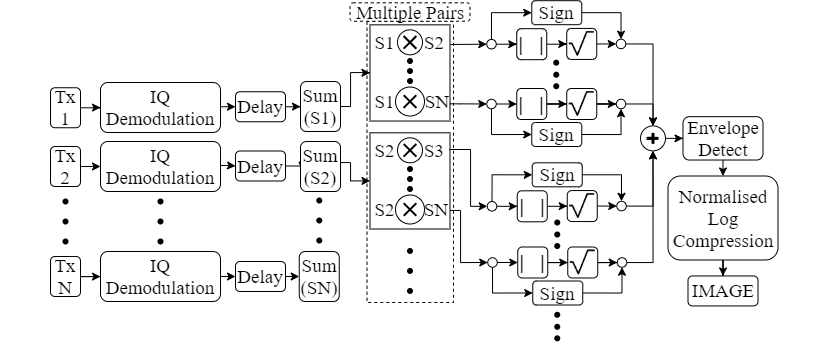}}
 \hfill
\subfloat[Row-Column specific Frame Multiply and Sum Pipeline \label{fig:RC_FMAS_pipeline}]{\includegraphics[width=\linewidth]{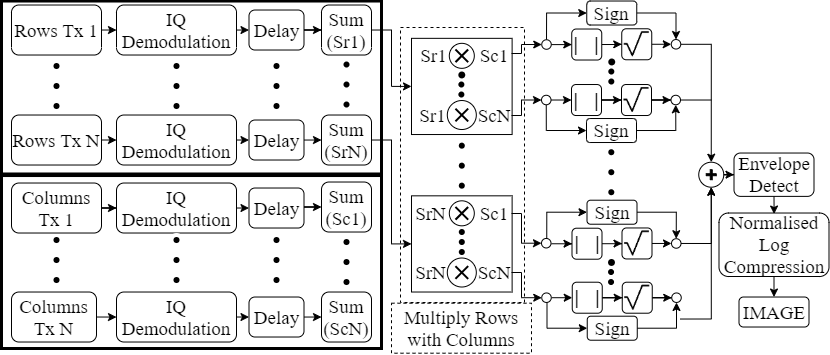}}
 \hfill
\caption{Pipelines for three different methods compared in this paper. Tx denotes a row transmission event which is then accompanied by a column reception or vice versa.}
\label{fig:Pipelines}
\end{figure}

The technique outlined above is the direct application of the FMAS to RCA. Here we  propose to perform the multiplication in a more tailored way to take advantage of the incoherent nature of row transmissions and column transmissions data. The~{\it{`cross-like'}} artefacts produced from row-column transmissions are produced when summing multiple~{\it{`line-like'}} PSFs. By performing FMAS, but only pairing row based transmissions with column based transmission, and not column with column or row with row the~{\it{`cross-like'}} artefacts can be greatly reduced, whilst keeping the noise reduction and other artefact suppression associated with current FMAS beamforming. Using this new technique the new number of pairs $N_P'$ becomes 

\begin{equation}
\label{eq:N-RCFMAS}
N_P' = N_{R_{tx}} \cdot N_{C_{tx}},
\end{equation}

where $N_{R_{tx}}$ and $N_{C_{tx}}$ are the number of rows transmissions and columns transmissions respectively. The new $RF_{FMAS}'$ will now be 

\begin{equation}
RF_{FMAS}' = \sum_{i= 1}^{N_{R_{Tx}}} \sum_{j=1}^{N_{C_{Tx}}} sign(RF_{ij}) \cdot \sqrt{\Big\vert RF_{ij}\Big\vert}.
\end{equation}
where $i$ now represents the row transmissions and $j$ represents the column transmissions.  Figure~\ref{fig:RC_FMAS_pipeline} shows gives a schematic representation of the new RCA specific FMAS pipeline.

\subsection{Simulations}
\begin{figure} [H]
    \centering

    \includegraphics[width=\linewidth]{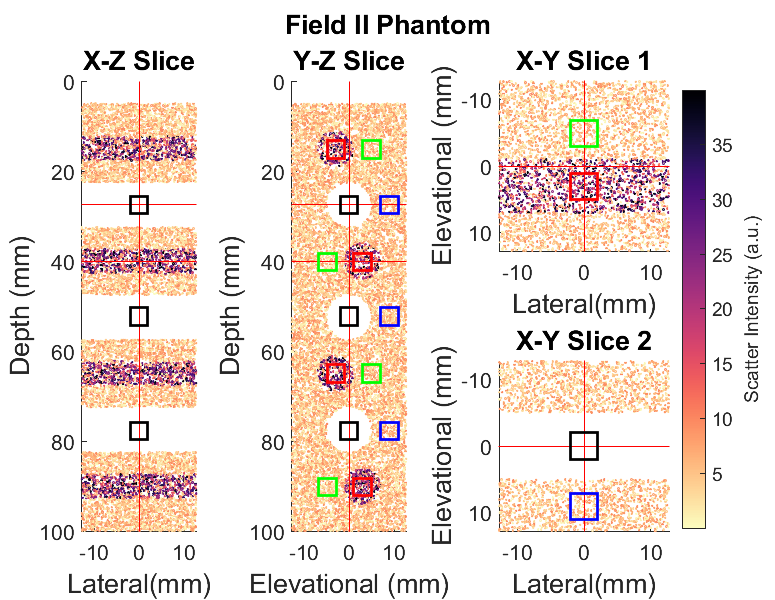}

    \caption{Cross sections of the Field-II phantom. The red lines represent the locations of the slices. These same slices are shown using the three methods, see Figure~\ref{fig:cyst_results}. The red and green squared represent the cyst and tissue regions respectively used for calculating TCR. The black and blue squares represent the noise and tissue regions respectively used for calculating the TNR. }
    \label{fig:phantom}
\end{figure}

Our study presented here was carried out exclusively  {\it{in silico}}. Simulations were conducted using the Field-II simulation environment~\cite{jensen_field_1996, jensen_calculation_1992}. The RCA was simulated by generating two overlapping transducers, one representing the columns, one representing the rows. Each transducer consisted of 128 elements and was place orthogonally to the other. The rows were defined to align with the x-axis and the columns were aligned with the y-axis. The pitch between each transducer was 0.2 mm and the length of each element was 25.6 mm. The transmission frequency ($f_0$) was 5 MHz and the bandwidth used was 6 MHz. These settings were chosen as it matched the transducer parameters of a transducer which will be used for future {\it{in vitro}} and {\it{in vivo}} verification. For row transmissions an impulse was given to the row elements and zero impulse was given to the column elements and then in reception only the column channels were utilised. For column transmissions the opposite was the case. The plane waves were produced by applying delays to either the rows or the columns thus allowing for steering either in the elevational or the lateral direction. For apodisation a Tukey apodisation of 0.5 was applied to the rows and columns in both transmission and reception. For beamforming an in-house beamformer was developed in the CUDA GPU programming environment~\cite{nvidia_and_vingelmann_peter_and_fitzek_frank_hp_cuda_2020}. Two types of simulations were performed during this study. The first was a simple single scatterer, the response of which was investigated to generate PSF of different techniques. The second simulation was a tissue mimicking simulation which consisted of a background of scatters along with a series of highly scattering tube regions and a series of empty tubes simulating cysts. Figure~\ref{fig:phantom} shows the tissue phantom used. 
White Gaussian noise was added with a signal to noise ratio of 20 dB to all simulations.

\vspace{-1cm}


\subsection{Image Quality Metrics}
The Full Width Half Maximum ($FWHM$) of the central peak was used to evaluate the resolution. Two metrics, the Peak-Intensity-Ratio ($PIR$) and the Peak-to-Maximum-Side-Lobe-Ratio ($PMSLR$), were used to evaluate the image contrast. To quantify the image quality of the cyst phantom simulations two further metrics were used: Tissue Contrast Ratio (TCR) and the Tissue-to-Noise Ratio (TNR).

The $FWHM$ was calculated separately in all three dimensions. The elevational and lateral $FWHM$ will be equivalent due to the symmetry of the simulations. The $PIR$ was calculated by taking the sum of the intensity within main peak bounded by the the $FWHM$ and dividing it by the intensity of the total volume encompassing the full PSF present within the field of view of the probe.
\begin{equation}
PIR = \frac{\displaystyle  \sum_{n=1}^{N_{peak}} I(\vv{\bf{r_n}})}{\displaystyle \sum_{n=1}^{N_{all}}I(\vv{\bf{r_n}})},
\end{equation}

where $N_{peak}$ represents all voxels within main peak bounded by $FWHM$, $N_{all}$ represents positions of all voxels in the image, and $I(\vv{\bf{r}})$ is the intensity of voxel located at position $\vv{\bf{r}}$.
This metric is non universal and will be dependent on the size of the bounding volume. However, by keeping the bounding volume size constant between simulations the metric can be used as a comparison of the contrast between different techniques. 

The $PMSLR$ was defined as the ratio of the amplitude of the main peak divided by the amplitude of maximum side lobe. This will give another measure on the contrast of the image and will be a universal metric. However, unlike the $PIR$, it will not take into consideration all the smaller side-lobes in the image.

The TCR is given by 

\begin{equation}
\label{eq:TCR}
 \mathrm{TCR = 20\; log_{10}\left(\dfrac{μ_{Tissue_1}}{μ_{Tissue_2}}\right)},
\end{equation}

where the $μ_{Tissue_1}$ is the mean intensity of region within highly scattering cyst and $μ_{Tissue_2}$ is the mean intensity of tissue region at same depth.

The TNR is given by
\begin{equation}
\mathrm{TNR = 20\; log_{10}\left(\dfrac{μ_{Tissue}}{μ_{Noise}}\right)},
\end{equation}

where $μ_{Noise}$ is the mean intensity of blank noise filled region.


\subsection{Optimisation}
To compare the three techniques, DAS, FMAS and Row-Column specific FMAS (RC-FMAS), a simulation based optimisation sweep was performing using different transmission parameters. A point scatterer was imaged at a depth of 50 mm. The PSFs generated from these simulations were then evaluated and compared. DAS, FMAS and RC-FMAS were all evaluated and compared over a range of imaging parameters using the $FWHM$, $PIR$ and $PMSLR$ metrics. The imaging parameters sweep consisted of number of angles from 6 to 30 and angle range from 5$\degree$ to 45$\degree$. This was done to investigate the angle ranges and number of transmission angles where this method operated most efficiently. For all tests the number of transmission angles and the ranges were kept the same for the rows and the columns as is traditionally done in row-column array imaging.

\section{Results}

\begin{figure} [H]
    \centering

    \includegraphics[width=\linewidth]{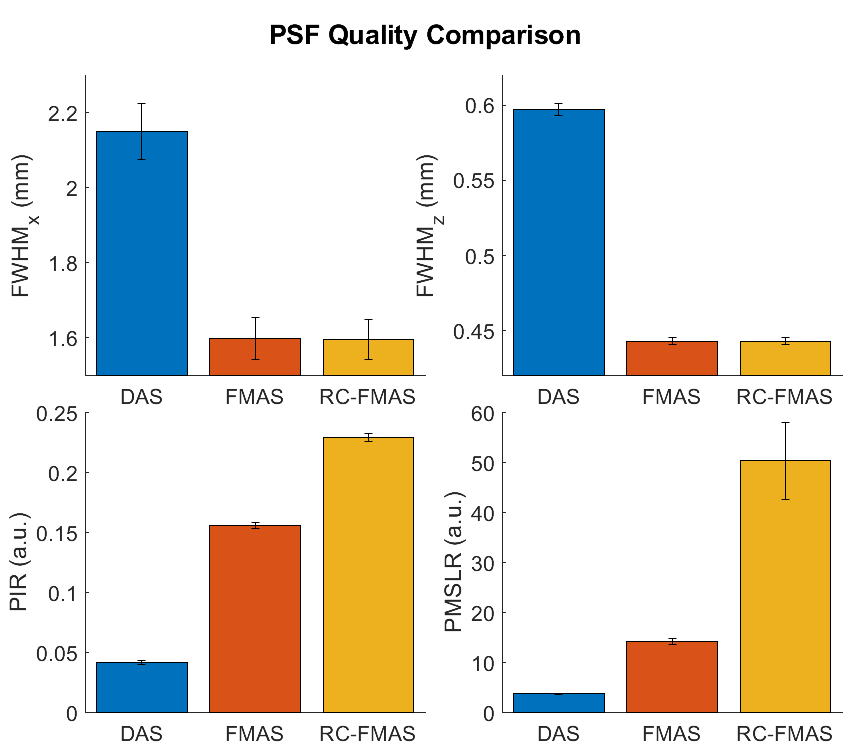}

    \caption{The PSF quality in high frame rate regime (Angle range: 10$\degree$, No. of Angles: 10). The $FWHM$ in x and y directions are identical so only one is shown. The values shown are the average of 100 scatterers located at depths from 1-10 cm. The errors shown are the standard error on the mean.}
    \label{fig:PSF_comparison}
\end{figure}

\begin{figure} [H]
    \centering

    \includegraphics[width=\linewidth]{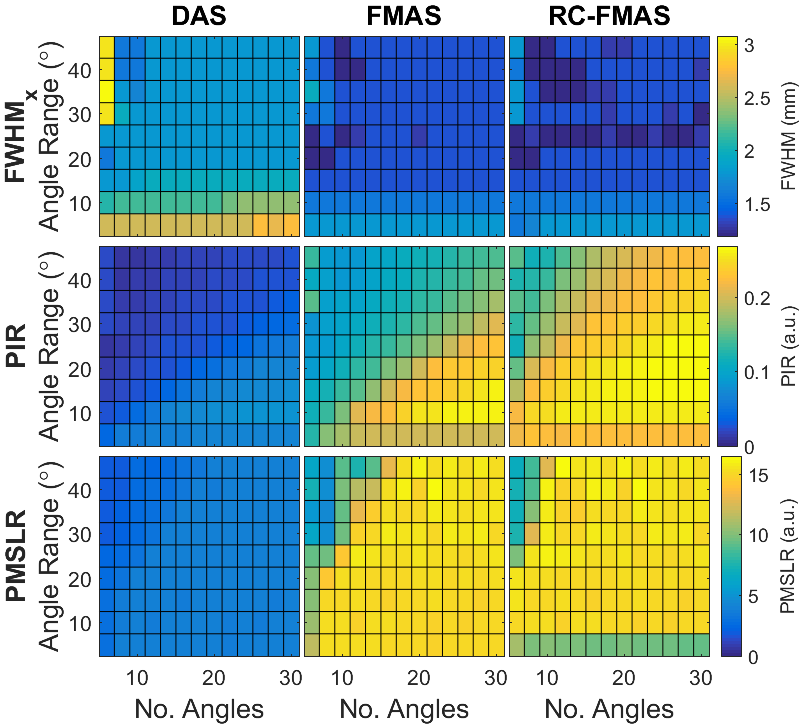}

    \caption{Results from the optimisation sweeps performed. The top row shows the $FWHM$ of the central point source aligned with the x direction. The middle row shows the $PIR$ and the bottom row shows the $PMLR$. For the $FWHM$ low numbers (blue) are show improved PSF resolution. For $PIR$ and $PMLSR$ higher numbers (yellow) shows reduction in imaging artefacts }
    \label{fig:Optimisation_sweep}
\end{figure}

The first test that was performed was the optimisation sweep. The results, as seen in Figure~\ref{fig:Optimisation_sweep}, show PSF improvement possible with both FMAS and RC-FMAS. In the high frame rate regime (Number of Angles: 10), when low angle ranges are used (Angle Range: 10$\degree$) , the $PIR$ is considerably higher going from 0.16 when using FMAS to 0.25 when using RC-FMAS.

\begin{figure} [H]
\centering
\includegraphics[width=\linewidth]{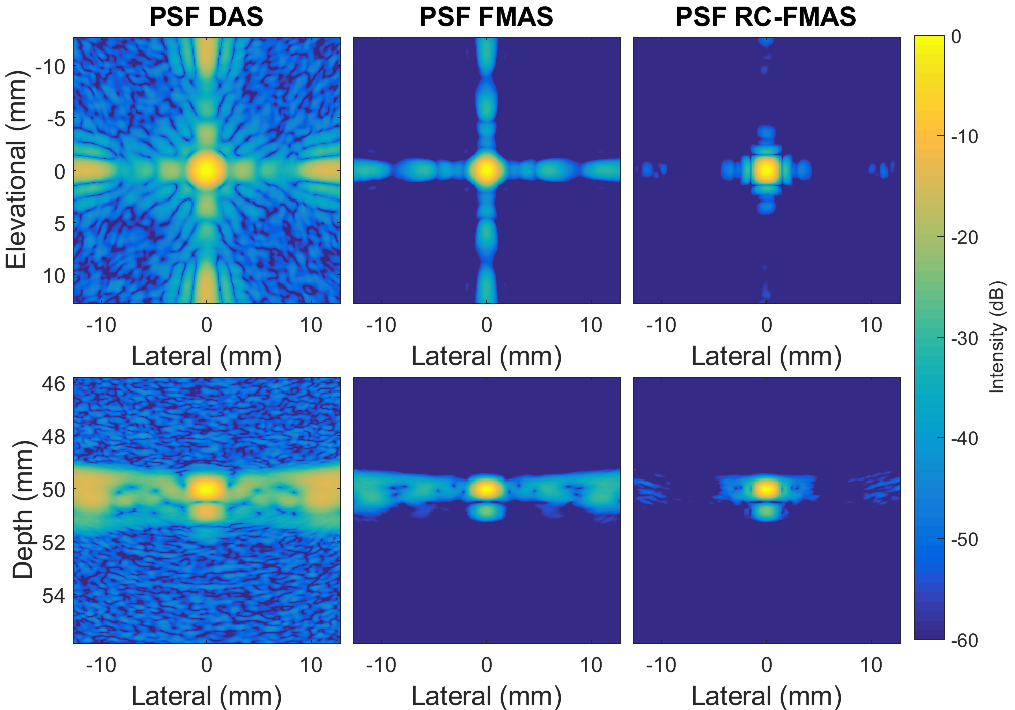}
\caption{Example PSF plots comparing the three methods, with angle range of 10$\degree$ and 10 transmission angles. This is an example of the improvement possible when few transmission angles are used. Top row shows x-y slice of the PSF, positioned at depth of 5 cm (location of point target). Bottom row shows x-z slice of the PSF, positioned at middle of transducer (location of point target).}
\label{fig:PSF}
    
\end{figure}

\begin{figure} [H]
\centering
\includegraphics[width=0.8\linewidth]{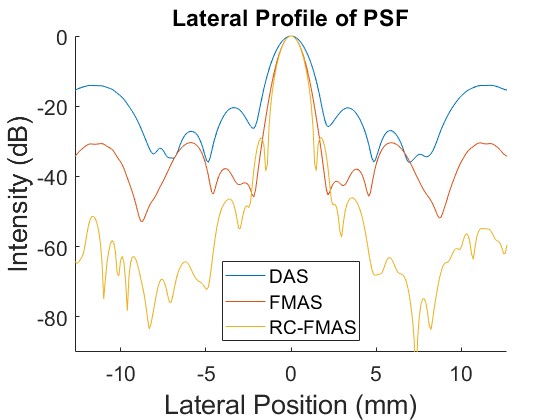}
\caption{Lateral Profile of the PSF with angle range of 10$\degree$ and 10 transmission angles.}
\label{fig:1D_PSF}
    
\end{figure}

To quantify the PSF improvement possible in this high frame rate regime, a point scatterer was imaged at multiple depths and the resulting image quality metrics where compared, see Figure~\ref{fig:PSF_comparison}. Figures~\ref{fig:PSF} and~\ref{fig:1D_PSF} demonstrate the PSF improvement possible using RC-FMAS when compared to either of the other two techniques.

\begin{figure}[H]
\centering

\subfloat{\includegraphics[width=\linewidth]{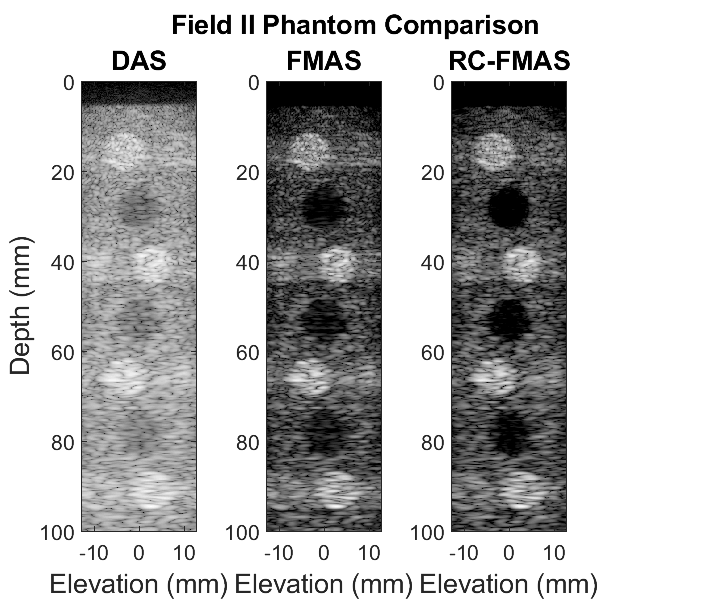}} \hfill\vspace{-1.4cm}
\subfloat{\includegraphics[width=\linewidth]{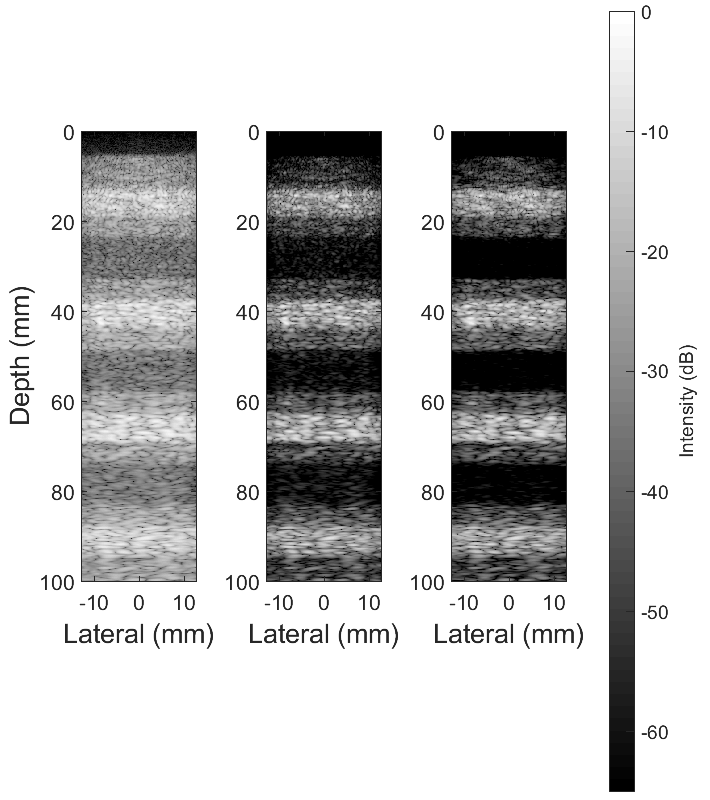}} \hfill\vspace{-1.85cm}
\subfloat{\includegraphics[width=\linewidth]{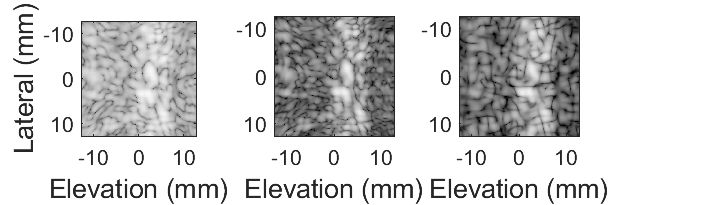}} \hfill\vspace{-0.16cm}
\subfloat{\includegraphics[width=\linewidth]{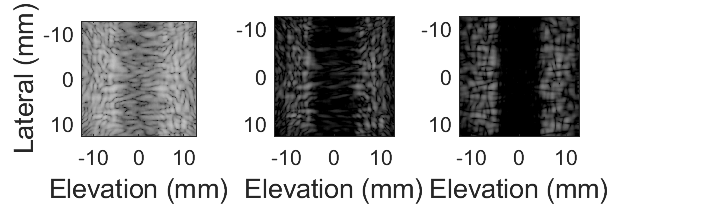}} \hfill
\caption{The cyst phantoms generated using Field-II. Top row is the standard DAS method, middle is the traditional FMAS method and the bottom row is the specific RC FMAS method. Locations of the slices can be seen in Figure~\ref{fig:phantom}.}
\label{fig:cyst_results}
\end{figure}

\begin{figure}[H]
\centering

\subfloat{\includegraphics[width=0.49\linewidth]{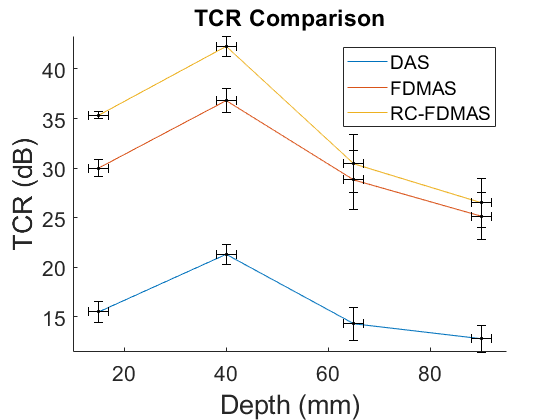}}
\subfloat{\includegraphics[width=0.49\linewidth]{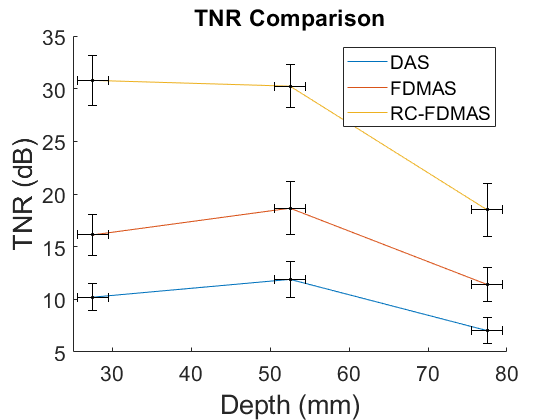}}

\caption{TCR and TNR values plotted at various depths. Regions used to calculate values shown can be seen in Figure~\ref{fig:phantom}. Each region was split into 4 sub regions and values were calculated separately then averaged, this was used to calculate the error in TCR and TNR. The error in depth is due to region used spanning a depth of $\sim$4 mm. }
\label{fig:TCRTNRRESULTS}
\end{figure}

Figure~\ref{fig:cyst_results} is a demonstration of the image quality improvement offered by the FMAS method and new RC specific FMAS method. 
The TCR and the TNR values were calculated for the three methods using this transmission scheme at different depths throughout the image. Figure~\ref{fig:TCRTNRRESULTS} shows the results obtained. As expected a general reduction in image quality is seen with depth. However, across all depths image quality increases when using FMAS and then increases further when using RC-FMAS when compared to DAS.

\vspace{-0.5cm}

\vspace{-0.8 cm}

\section{Discussion}
By applying FMAS to RCA probe,  artefact reduction and noise suppression was demonstrated with the TCR and TNR improving by up to $\sim$16dB and $\sim$6dB respectively. By amending the FMAS method, taking advantage of the intrinsic lack of coherence between row and column data, TCR and TNR improved by a further 6 dB and  15 dB  respectively. A 56\% saving in computational cost was found between FMAS and RC-FMAS (when 10 transmission events are used). An optimisation sweep showed major artefact reduction during all acquisition schemes tested. 

From Figure~\ref{fig:Optimisation_sweep} the general shape of the optimisation space did not change between the three methods, but a clear improvement is seen when comparing DAS to FMAS and then again from FMAS to RC-FMAS, particularly when considering the $PIR$ (note higher $PIR$ and $PMSLR$, indicates improved image quality potential).  The FMAS method leads to major improvements in $FWHM$ in lateral and elevational directions, with RC-FMAS then providing further minor improvements. The axial $FWHM$ showed minor improvement with FMAS and RC-FMAS methods. The variance in $FWHM$ demonstrated is caused by side lobes merging with main lobe. From Figure~\ref{fig:Optimisation_sweep} a general sense of the optimum imaging parameters can be determined. The $PIR$ and $PMSLR$ improved significantly using both methods. As expected the extremes (high angle range with low number of angles and low angle range) lead to a reduction in PSF quality. Comparing the $PMSLR$ in the extreme low angle range can be non representative. As $PMSLR$ is dependent on the peak side lobe intensity, side-lobes at different lateral positions can be compared. In the case of 5$\degree$ angle range the $PMSLR$ in the RC-FMAS case gives a worse value. However, this is due to additional intermediate side-lobes appearing close to the main lobe see Figure~\ref{fig:1D_PSF}. This anomaly will be produced by the multiplication of transmissions, and occurs for both RC-FMAS and FMAS. But in this particular case, for FMAS, they are covered by the main lobe.

For most applications, imaging will be performed in the centre of the optimisation sweep. This region is useful as the frame rate is typically sufficiently low whilst image quality can be maintained. 

The most significant improvement was seen in the reduction of distant side-lobes as shown in Figure~\ref{fig:1D_PSF}.
The PSF quality will be dependent on the depth. When imaging deeper points angular range cannot be too high otherwise transmitted waves will no longer overlap.
The results of the cyst phantom experiment, see Figure~\ref{fig:cyst_results} demonstrate clearly that the contrast is greatly improved using both the FMAS method and RC-FMAS method, along with significant noise reduction. Quantitative improvement is also shown with significant improvement in both TCR and the TNR when using the RC-FMAS versus any of the other two methods, as outlined above.

The major advantage of this method is that it allows for higher quality imaging at an increased frame rate. Using less acquisition angles will reduce motion artefacts and and aids in the imaging of fast moving vessels. This will be of particular use when performing cardiac imaging. It also will facilitate faster acquisitions of super resolution images, which coupled with artefact reduction should lead to better quality super resolution images with RCA probes.

A disadvantage of this method is slightly increased computational cost when beamforming. The total number of operations has increased from order $N$ to order $N^{2}$. As RCA probes use a reduced number of channels compared to most other 3D imaging techniques and because of the reduced number of transmission angles needed for this method the final computational cost will not be a significant challenge particularly with increase in computational power offered by modern graphical processing units. Although all simulations were performed as realistically as possible  this is an {\it{in silico}} experiment and therefore limited. Consequently, {\it{in vitro}} validation and {\it{in vivo}} demonstrations are both required. 

Improving the image quality of RCA probes will allow for them to potentially be applied across a range of medical imaging applications, due to it's ability to produce low cost high quality 3D ultrafast imaging. Cardiovascular imaging, Super resolution and cancer imaging/screening are all potential application areas of particular interest.


\section{Conclusion}
In this paper two new methods for generating images using a row-column array has been proposed and tested. The first simple FMAS method is an existing method which has been applied to the RCA probe for the first time. The second method was a new technique adapted from the FMAS specifically tailored for RCA probes. Image quality improvement of up to  $\sim$20dB was seen when using both methods compared to DAS. Further image improvement was then seen when using the RC-FMAS method compared with the traditional FMAS method. The reduction in side lobe artefacts has led to a clear improvement in B-mode images. In the future these methods will need to be tested in vivo and in vitro. They will then hopefully lead to significant improvements in all types of RCA imaging for high quality 3D ultrafast ultrasound imaging.  

\section{Acknowledgements}
We would like to acknowledge the funding from the EPSRC CDT in Smart Medical Imaging, the EPSRC project grant (EP/T008970/1) and the Department of Bioengineering at Imperial College London. Dr Chee Hau Leow is also thanked for his contribution to the GPU beamforming code.


\printbibliography

\end{document}